\documentclass[12pt,preprint]{aastex}

\shorttitle{Stability of Cylindrical Tangential Discontinuity}
\shortauthors{Ershkovich and Israelevich}

\begin{document}


\title{On the Stability of Cylindrical Tangential Discontinuity, Generation and Damping of Helical Waves}


\author{A. I. Ershkovich and P.L. Israelevich}
\affil{Department of Geophysics and Planetary Sciences, Faculty of Exact Sciences, Tel Aviv University, Tel Aviv, 69978, Israel}



\begin{abstract}
Stability of cylindrical interface between two ideal incompressible fluids, including the magnetic field, surface tension and gravitational field is studied in linear approximation. We found that helical waves arising both in plasma comet tails and on the vertical cylindrical water jet in the air are described by the same dispersion equation where the comet tail magnetic field plays the same stabilizing role as surface tension for water jet. Hence they represent the same phenomenon of Kelvin-Helmholtz instability. Thus, helical waves in planetary and cometary magnetotails as well as in astrophysical jets may be simulated in the laboratory. The resonance nature of the instability damping is demonstrated.
\end{abstract}



\section{Introduction}

Stability of plane interface between two ideal incompressible fluids has been first considered by \citet{Kelvin1871}  (see also \citep{Landau1959, Milne1960}).

Small oscillations in such a fluid are always potential in the first approximation. Therefore, the velocity potential satisfies the Laplace equation, and for perturbations proportional to $\exp[i(kx-\omega t)]$  we arrive at the dispersion equation 
\begin{equation}
\frac{\omega }{k}=\frac{\rho _{1}V _{1}+\rho _{2}V _{2}}{\rho _{1}+\rho _{2}}\pm \left [ -\frac{\rho _{1}\rho _{2}\left ( V_1-V_2 \right )^2}{\left ( \rho _{1}+\rho _{2} \right )^2}+\frac{\rho _{1}-\rho _{2}}{\rho _{1}+\rho _{2}}\left ( \frac{g_n}{k} \right )+\frac{\sigma  k}{\rho _{1}+\rho _{2}} \right ]^{1/2},
\end{equation}
where $\rho$ is the density, $V$ is the velocity, $g_n$ is the normal (to the interface) component of the gravitational acceleration, and $\sigma$ is the surface tension. \citet{Rayleigh1892} was the first to consider a cylindrical interface but his stability analysis was restricted by "varicose" perturbations (in modern nomenclature, sausage-like ones) also proportional to $\exp[i(kz-\omega t)]$, and hence could not  describe kink-modes proportional to $\exp[i(kz+m\varphi-\omega t)]$, $m\neq 0$.

More general analysis was required following a discovery of the geomagnetic tail by \citet{Ness1965}. It has been performed in \citep{McKenzie1970} for a cylindrical interface in compressible plasma in MHD approximation (the dispersion equation in this case is transcendental and may be solved only numerically), and in \citep{Ershkovich1971,Ershkovich1972} for incompressible plasma with the magnetic field (like the plasma bulk velocity) parallel to the cylinder axis $z$. The solution of Laplace equation proportional to $\exp[i(kz+m\varphi-\omega t)]$  describes helical waves, like kink modes $m = 1, 2$, etc., along with sausage mode $m = 0$ (the solution is single-valued only if $m$ is an integer, positive or negative). If such solutions are possible then helical waves may exist in the nature, identified and observed.

Indeed, in contrast to the Earth magnetic tail, such oscillations have long been observed {\it visually} in rectilinear comet plasma tails (type I tails). \citet{Bessel1836} was the first to describe these wave motions in detail. \citet{Alfven1957} assumed that they are MHD waves. But helical waves arising due to interface instability are surface waves propagating in both fluids (as a whole). Hence, they cannot be Alfv\'{e}n waves as the Alfv\'{e}n velocity in the comet tails (where the plasma is heavy) is usually much less than in the neighboring solar wind. The quantitative MHD theory of helical wave origin in type I comet tails due to instability in plasma cylinder was suggested in \citep{Ershkovich1972a,Ershkovich1973,Ershkovich1980}. The corresponding dispersion equation for the model of plasma cylinder with radius $R$ is 
\begin{equation}
\frac{\omega }{k}=\frac{\rho _{i}V _{i}-L_m\rho _{e}V _{e}}{\rho _{i}-L_m\rho _{e}}\pm \left [ \frac{B^2_i-L_mB^2_e}{4\pi \left ( \rho _i-L_m \rho _e \right )}+\frac{L_m \rho _{i}\rho _{e}\left ( V_e-V_i \right )^2}{\left ( \rho _{i}-L_m \rho _{e} \right )^2} \right ]^{1/2},
\end{equation}
where indices $i$ and $e$ refer to internal and external plasmas, respectively, and the function
$L_m=[I'_m(kr)K_m(kr)]/[I_m(kr)K'_m(kr)]$  is taken at the unperturbed interface $r = R$. 
$I_m$ and $K_m$ are modified Bessel functions, a stroke denotes the derivative by argument $kr$. The function $L_m (kR)$ is always negative: $0 \geq L_m \geq -1$ (see Figure 1). 

By using the same standard procedure of linearization described in more details in \citep{Ershkovich1972,Ershkovich1980} we may also include the gravitational field ${\mathbf g}$ and surface tension $\sigma$, arriving at the following dispersion equations: for cylindrical interface between two liquids
\begin{equation}
\frac{\omega }{k}=\frac{\rho _{i}V _{i}-L_m\rho _{e}V _{e}}{\rho _{i}-L_m\rho _{e}}\pm \left [ \frac{B^2_i-L_mB^2_e}{4\pi \left ( \rho _i-L_m \rho _e \right )}+\frac{L_m \rho _{i}\rho _{e}\left ( V_e-V_i \right )^2}{\left ( \rho _{i}-L_m \rho _{e} \right )^2} +\frac{g_n}{k} \frac{\left ( \rho _{i}+L_m \rho _{e} \right )}{\left ( \rho _{i}-L_m \rho _{e} \right )}+\frac{\sigma  k}{\rho _{i}-L_m\rho _{e}} \right ]^{1/2},
\end{equation}
and for liquid-gas interface 
\begin{equation}
\frac{\omega }{k}=\frac{\rho _{i}V _{i}-L_m\rho _{e}V _{e}}{\rho _{i}-L_m\rho _{e}}\pm \left [ \frac{B^2_i-L_mB^2_e}{4\pi \left ( \rho _i-L_m \rho _e \right )}+\frac{L_m \rho _{i}\rho _{e}\left ( V_e-V_i \right )^2}{\left ( \rho _{i}-L_m \rho _{e} \right )^2} +\frac{g_n}{k} \frac{\left ( \rho _{i}+L_m \rho _{e} \right )}{\left ( \rho _{i}-L_m \rho _{e} \right )}+\frac{\sigma  k}{\rho _{i}-L_m\rho _{e}} \left ( \frac{I'_m}{I_m} \right ) \right ]^{1/2}.
\end{equation}
The difference in the last term between equations above is due to the fact that $\sigma = 0$ for the gas which is located outside the cylinder ($r > R$).

\section{Discussion}

With $V_e = V_i = V$, $g_n = 0$, $\sigma  = 0$ equations (3) and (4) describe stable "surface Alfv\'{e}n waves" convected with the fluid bulk velocity $V$: $\omega /k=V \pm [(B^2_i-L_mB^2_e)/(4\pi \rho _i - 4\pi L_m \rho _e)]^{1/2}$. Standard expression $V_A = B /\sqrt{4 \pi \rho}$  is obtained from the expression above with $\rho = \rho _i \gg \rho _e$, $B = B_i \gg B_e$.

In the short wavelength limit $kR \gg 1$, $L_m \rightarrow  -1$, and with $g = 0$, $\sigma  = 0$ equation (3) reduces to the dispersion equation for plane interface obtained by \citet{Syrovatskii1953} within MHD for incompressible plasma. With $kR \gg 1$ and $B = 0$ equation (3), naturally, reduces to the dispersion equation (1). Finally, with $B = 0$, $g_n = 0$, $V_e = 0$ equation (4) is the solution of the dispersion equation (2.1) in \citep{Yarin2011} for a vertical cylindrical water jet in air. Strictly speaking, the dispersion equation (2.1)  for water jet \citep{Yarin2011} was obtained for perturbations proportional to  $\exp[i(kz-\omega t)]$. Such perturbations are plane waves, which, in principle, cannot correspond to helical waves. But it is clear that the structure of the dispersion equation for cylindrical jet, being dependent on combinations  $I_m$, $K_m$ and their first derivatives, has the same form for all integer values of $m$. 

As $L_m < 0$, $I'_{m}/I_m > 0$ both magnetic field and surface tension terms are always positive and hence tend to stabilize the interface. The velocity shear term under the radical is negative being responsible for the Kelvin-Helmholtz instability. The term proportional to $g_n$ results in the flute (Rayleigh-Taylor) instability for heavy fluid over the light one, and with $\rho _i \gg \rho _e$, $V = 0$ it describes stable gravitational waves on deep water: $\omega ^2 = kg$. 

External gravitational field always tends to violate the cylindrical symmetry, because the external normal to the cylindrical surface changes its sign (relative to the field) at the opposite sides of the cylinder. 

There are, however, several cases when $g_n$ is small or even vanishes. This is, of course, almost rectilinear (and cylindrical) plasma tail of comets, where helical waves are visually observed (see e.g. \citep{Ershkovich1980} and references therein) as well as astrophysical jets, e.g. \citep{Birkinshaw1996}.  Formation of plasma comet tails is governed by solar wind, and, hence, they are almost in antisolar direction, with rather small aberration angle due to orbital motion around the Sun. Another example (although hypothetical) is self-gravitating astrophysical jet where the gravitational field is radial and, hence, axially symmetric. 

Stability of vertical water jets has been studied in  laboratories for a long time (see, e.g., \citep{Yarin2011,Leibovich1983,Gallaire2003}. 
Let us now estimate the minimal velocity shear $V_{min}(R)$ required for instability of vertical water jet in air. With $\rho _i \gg \rho _e$, $B = 0$, $V_e = 0$ (air is at rest), $g_n = 0$, $V = V_i$, equation (4) yields
\begin{equation}
\frac{\omega }{k}=V \pm \frac{1}{\sqrt{\rho _i}}\sqrt{L_m \rho _e V^2+\sigma k\frac{I'_m}{I_m}},
\end{equation}
whence one obtains the instability criterion in the form
\begin{equation}
|L_m| \rho _e V^2 > \sigma k \frac{I'_m}{I_m}.
\end{equation}

As functions $L_m$  and $I'_{m}/I_m$ depend on $kR$, the value $kR$ is to be estimated.

Observations in plasma comet tails show that $kR \sim 1$ (e.g. \citep{Ershkovich1980}, Table II). This fact is not unexpected as the cylinder radius $R$ (more precisely, the circumference $2\pi R$) is the only characteristic length scale of the problem under consideration. Observations in water jet are also indicative of $kR \sim 1$, and the first attempt to explain this phenomenon was performed by \citet{Rayleigh1892}. It is indeed tempting to explain this fact by means of behavior of the instability growth rate, $\gamma = \mathrm{Im } \omega$. But this is not so: the function $\gamma (kR)$ has no maximum with $kR = 1$. According to equation (5), 
\begin{equation}
\gamma = \frac{k}{\sqrt{\rho _i}}\sqrt{|L_m| \rho _e V^2-\sigma k\frac{I'_m}{I_m}}.
\end{equation}

Hence, $\gamma \propto k \sqrt{|L_m|}$   if $\sigma \rightarrow 0$ or the velocity of jet, $V$, is large enough. The factor $ x\sqrt{|L_m(x)|}$   is shown in Figure 2. It is seen that the dependence of the instability growth rate $\mathrm {Im }\omega = \gamma (kR)$ is monotonous, and has no maximum with $kR \sim 1$.  

Fluid parameters are varying in broad ranges while $kR \sim 1$ remains almost the same. It means that $kR \sim 1$ is a geometric characteristic, a peculiarity of a cylinder. Observation mentioned above seems not to be explained within a framework of infinitesimal amplitudes. On the other hand, a finite amplitude theory \citep{Ershkovich1973} shows that the critical wave amplitude $\delta _c$ of non-linear stabilization, indeed reaches maximum with $kR \sim 1$ as $\delta _c /R = (\Lambda _1)^{-1/2}$, and the function $\Lambda _{1} (kR)$  has here the sharp minimum (Figure 4 in \citep{Ershkovich1980} shows the function $\Lambda _{m} (kR)$  for $m =0$ and $m = 1$).

According to Laplace formula, an additional surface tension under the cylinder interface is $\sigma /R$, so that with $kR = 1$ the coefficient  $\sigma =$  74 dyn cm$^{-1}$ (for a plane water surface in air) is double valued, and using equation (7) we arrive at the conclusion that the instability arises if $V^2 > (2 \sigma I'_m/I_m)/(\rho _e R |L_m|).$ 

Finally, with $\sigma  = 74$ dyn cm$^{-1}$, $\rho _i /\rho _e = 770$ for the water-air interface, $|L_1| \approx 0.75$ and $I' _{m}/I_{m} \approx 1.25$ for kink mode $m = \pm 1 $ (\citep{Kruskal1958}, Figure 1) one obtains    
\begin{equation}
V_{min}[\mathrm{m  s^{-1}}] = \sqrt{\frac{20}{R\mathrm{[cm]}}}.
\end{equation}

Thus, the minimal initial jet velocity of water vertical free fall required for instability is 4.5 m s$^{-1}$, 2.0 m s$^{-1}$, and 1.4 m s$^{-1}$ for the water jet radius $R = 1,$ 5, and 10 cm, respectively. The value $V_{min}$ happens to be too high in order to observe helical waves in the cylindrical vertical jet from a water tap, but, instead, it is possible to observe there stable surface waves (i.e. normal modes of oscillations) traveling along the jet both upstream and downstream (in the frame of reference moving with the velocity $V$) when the expression under the radical in the equation (5) is positive. This phenomenon is just the same as ripples created by wind on the lake surface. 

Helical waves in plasma comet tails may become visible only when their amplitude becomes large enough. As a result, this phenomenon is observed relatively seldom. The stability conditions seem to be marginal. As helical waves in comet tail and in vertical water jet obey almost the same dispersion equation this astrophysical phenomenon may be (and, in our opinion, is to be) simulated in laboratory.

We did not consider here the effect of finite width, $d$, of a transition layer between two fluids. It is known to be small with $kd \ll 1$ ,  and as for cylindrical jet $kR \sim 1$, this effect is negligible if $d \ll R$ which is the case for water jet in air and seems to be observed (as sharp decrease of brightness) for plasma comet tails. Model of cylindrical comet tail with transition layer of finite thickness $d$ is treated by \citet{Chen1982}.

\section{Resonance damping of helical waves}

\citet{Landau1944} found a sharp decrease of the Kelvin-Helmnoltz instability growth rate when the phase velocity of surface wave, $\mathrm {Re } \omega /k$  is approaching the acoustic velocity, $c$, with full damping  $\gamma = \mathrm {Im } \omega  = 0$ when the phase velocity reaches $\sqrt{2}c$  (see also \citep{Landau1959}, ch.9, $\S$  84). A similar effect was  described in \citep{Ray1983} for magnetoacoustic velocity. We believe that there is a simple explanation: when the phase velocity of surface wave, $\mathrm {Re } \omega /k$, approaches the characteristic velocity of normal mode of oscillations in the fluid, a resonance arises, and stable hydrodynamical or MHD modes are generated in the whole fluid volume. But the Kelvin-Helmholtz instability of the tangential discontinuity is a surface phenomenon, with the amplitude of perturbation decreasing (in the plane case - exponentially) away from the interface. Thus, these waves, in some meaning, are two-dimensional, with relatively restricted stock of kinetic energy, supplied by the velocity shear. When this energy is transferred from 2D to 3D space generating stable waves everywhere, the energy stock is rapidly exhausted, and instability is damping.  But if so, the same phenomenon should exist in incompressible plasma while reaching the Alfv\'{e}n velocity, $\omega / k \approx V_A$. Indeed, this effect was described in \citep{Ray1983, Lau1981}for plane interface.

Let us consider a cylindrical plasma jet (with the velocity $V_i$) immersed into plasma at rest, with the same parameters, i.e. $\rho _i = \rho _e = \rho$, $B_i = B_e = B$, and $V_e = 0$ (alternatively, we may choose the frame of reference where $V_e = 0$). Then equation (2) yields the phase velocity $\mathrm{Re} \omega /k = V/(1-L_m)$, where $V = V_i$ is the velocity jump. The radical in equation (2) vanishes, and the interface becomes stable with $V=(1-L_m)V_A/|L_m|$, and the phase velocity $\omega /k = V/(1-L_m)=V_A/|L_m|=1.15 V_A$  for kink mode $m =1, L_m = -0.75$. For plane interface $L_m = -1$, hence $V = 2V_A$, and $\omega /k = V_A$.

A possibility of resonance generation of Alfv\'{e}n waves in the whole volume of fluid with $\omega \rightarrow kV_A$  seems to be obvious. The fact that the magnetized shear layer is stable if its Alfv\'{e}n speed is greater than half the velocity jump across the interface was found in \citep{Ray1983, Lau1981} (unfortunately, the resonance nature of the instability damping has not been mentioned therein). According to \citet{Ray1983}, the interface remains stable (despite the growing flow velocity $V$) when $V \geq 2c$ and $V_A \geq c$.  As the phase velocity $\omega /k = V/2$ , these conditions may be rewritten in the form $\omega /k \geq c \leq V_A$ , which agrees with the resonance scenario above: the instability ceases because the energy supplied by the velocity shear transfers (due to the resonance) for excitation of normal modes of the fluid oscillation, first, of sound waves (as $c \leq V_A$), and then, of MHD waves (Alfv\'{e}n and magnetosonic).

Similar resonance damping occurs with unstable capillary waves. In order to demonstrate this effect, let us assume that the liquid in a cylindrical jet (moving with the speed $V$) has almost the same density, $\rho _i$, as the ambient liquid at rest, i.e. $|\rho _i - \rho _e| \ll \rho _{i,e} = \rho$. Then equation (3), with $B =0$, $g_n =0$ yields $$\frac{\omega }{k}=\frac{V}{1-L_m}\pm \left [ \frac{L_m V^2}{\left ( 1-L_m \right )^2}+\frac{\sigma  k}{\rho (1 -L_m)} \right ]^{1/2}.$$ The radical vanishes if $$V=\left ( \frac{1-L_m}{|L_m|} \right )^{1/2}\sqrt{\frac{\sigma k}{\rho }},$$ and the phase velocity $$\frac{\omega }{k}=\frac{V}{1-L_m} =\sqrt{\frac{\sigma  k}{\rho |L_m| (1 -L_m)}}.$$ Hence, for a kink mode $m =1$, $L_m = -0.75$ one obtains $\omega /k =0.87 \sqrt{\sigma k /\rho}$. For the plane interface $L_m = -1$ we find $\omega /k = \sqrt{\sigma k /(2 \rho)}$   (which, naturally, may be obtained directly from equation(1)). The classical value for stable capillary waves is $\omega /k =\sqrt{\sigma k /\rho}$ \citep{Kelvin1871, Landau1959}.

A small region of the cylindrical interface may be considered as plane for perturbations with $kR \gg 1$, and for plane case the dispersion equation for perturbations $\sim \exp[i(\mathbf{kr}-\omega t)]$  depends on scalar  products $\mathbf{kV}$ and $\mathbf{kB}$. This means that there are always directions along which the stabilizing role of the magnetic field becomes negligible. As short wavelength perturbations may propagate in all directions a tangential discontinuity always remains unstable. But this is not the case for helical waves propagating along the cylinder axis.

\section{Conclusion}

Dispersion equations (3) and (4) describe rather broad class of hydrodynamical and MHD instabilities and normal modes of oscillations of the cylindrical interface between two fluids, started with Alfv\'{e}n waves and gravitational waves on deep water to flute and Kelvin-Helmholtz instabilities in planetary and comet tails and water jets in air (including also capillar instability in liquids). Although they were obtained in linear approximation (and hence each of these effects may be studied independently) the stability criterion $\mathrm{Im }\omega  = 0$ depends on the balance of {\it all} the relevant terms under the radical. This balance is particularly important under marginal stability conditions when only their sum is indicative of stability or instability of the interface.

We also drew attention to the fact that the instability growth rate obtained in linear approximation cannot explain the preferential generation of modes with $kR \sim 1$. In particular, this fact refers to helical waves observed visually in comet plasma tails. At the same time, finite amplitude treatment \citep{Ershkovich1973,Ershkovich1980} seemed to explain these observations.

Both Alfv\'{e}n and capillary waves arising due to Kelvin-Helmholtz instability on the cylindrical interface have been considered. We arrived at the conclusion that sharp damping of these helical waves occurs when their phase velocity approaches the characteristic velocity of normal modes of oscillation, so that it has resonance nature.

Finally, we found that helical waves both in plasma comet tails and in vertical cylindrical water jet in the air are governed by almost the same dispersion equation (which means that, in fact, we deal with the same phenomenon). This fact allows us to suggest an idea of laboratory simulation of helical wave generation in cometary and planetary magnetotails as well as in astrophysical jets by using vertical water (or any other suitable liquid) jet.

\acknowledgments


\clearpage



\begin{figure}
\epsscale{.80}
\plotone{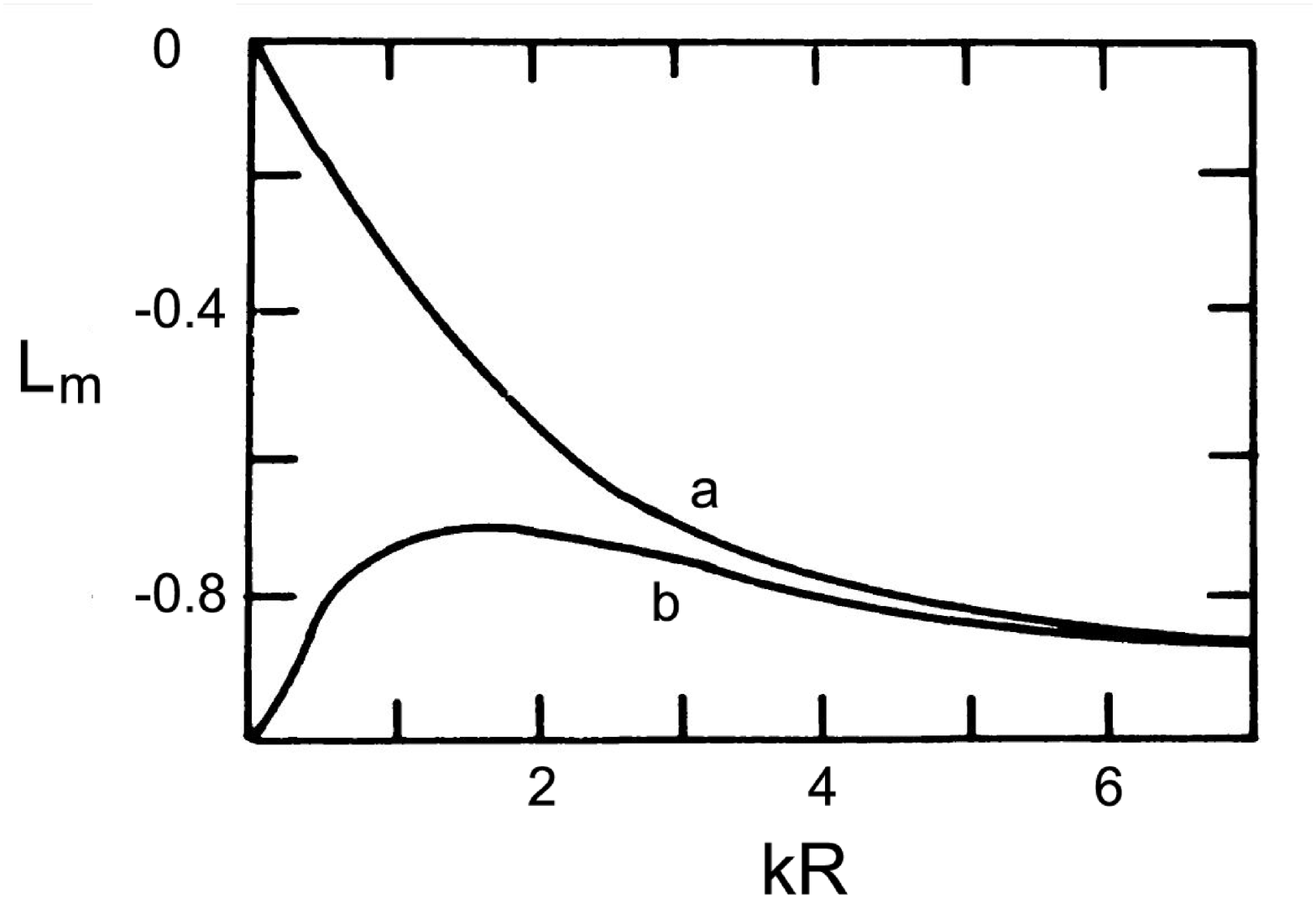}
\caption{The function $L_m(kR)$ with $m=0$ and $m=1$ (curves $a$ and $b$, respectively, according to \citet{Ershkovich1973}).\label{fig1}}
\end{figure}

\clearpage


\begin{figure}
\plotone{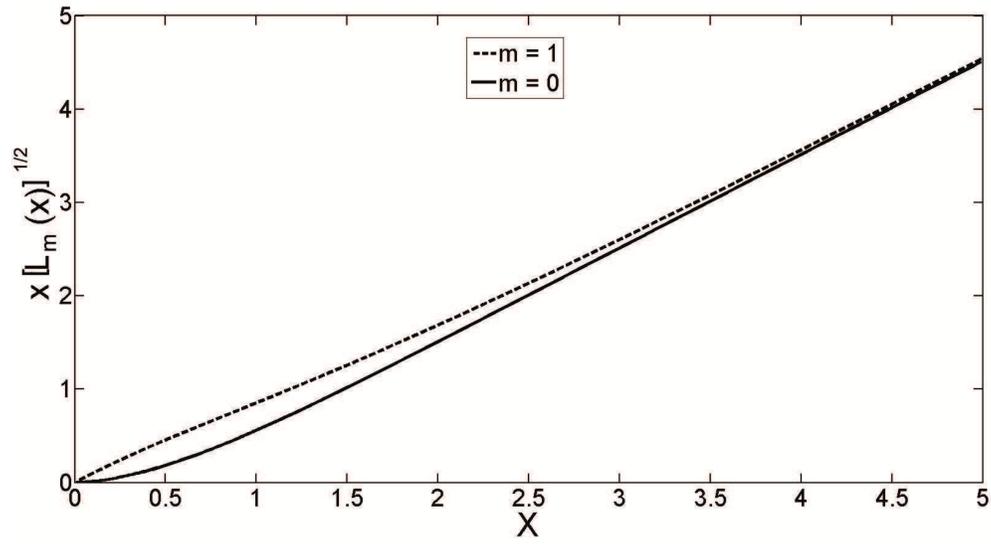}
\caption{The function $ x\sqrt{|L_m(x)|}$ with $m=0$ (solid line) and $m=1$ (dashed line), $x=kR$.\label{fig2}}
\end{figure}

\end{document}